# Vortex solitons in star networks

**Yaroslav V. Kartashov**
*Institute of Spectroscopy, Russian Academy of Sciences, Troitsk, Moscow, 108840, Russia*

I consider vortex solitons in two-dimensional star networks created by several one-dimensional line waveguide arrays (or rays) brought in close proximity. First waveguide in each array belongs to the ring of certain radius and its modal field overlaps with modal fields of waveguides from other rays laying on this ring, as well as with modal fields of waveguides in this line array. This creates a system with discrete rotational symmetry that is capable of supporting various localized linear vortex states due to local coupling between waveguides on the central ring, where line arrays approach each other. Localized vortex states emerge from the continuum of the states extended along the rays of the structure upon variation of the radius of the central ring. I show how different vortex modes with progressively increasing topological charges emerge when the number of rays in structure and, therefore, its discrete rotational symmetry increases. Each such linear localized mode gives rise to a family of thresholdless vortex solitons in nonlinear optical medium, whose topological charges are limited by the discrete rotational symmetry of the system. Vortex solitons in star networks exhibit qualitatively different behavior with increasing power depending on whether the propagation constant of linear vortex that gives rise to a family of solitons is located above or below the band of extended states. Stability properties of vortex solitons nontrivially depend on their topological charge and discrete rotational symmetry of the system, showing very different picture from usual ring-like waveguide arrays.

## 1. Introduction

Vortices are ubiquitous in nature and their formation is manifested in many physical phenomena observed in everyday life and in scientific laboratories. Vortices attract considerable attention and are actively studied in different areas of physics, including hydrodynamics, physics of plasmas, acoustics, electronic and optoelectronic systems, physics of matter waves, and optics. Vortices emerging in electromagnetic fields are of particular interest, since they promise multiple practical applications, ranging from information encoding, power transfer, trapping applications, transfer of angular momentum between light and matter, and formation of long-living propagating excitations, see reviews [1-6]. In nonlinear optical materials, vortex-carrying beams can exist as self-sustained excitations that do not diffract in the course of propagation [2,7-13]. However, if the nonlinear material is uniform, such self-sustained excitations appear to be very sensitive to azimuthal modulations are can be destroyed upon propagation due to azimuthal instabilities [14], unless competing, nonlocal, nonlinearities, or dissipative or some other stabilization mechanisms are involved, as described in reviews [11,12].

Among the most fruitful mechanisms of stabilization of self-sustained states carrying vorticity is the utilization of the optical potentials, i.e. transverse modulations of the refractive index of the optical material. Such localized or periodic (lattice) potentials profoundly improve stability of vortex solitons, see reviews [11,12,15], as well as experimental realizations in periodic lattices [16-20], and numerous theoretical results [21-26] on properties of discrete and lattice vortex solitons. Vortex solitons have been also actively studied in aperiodic lattices that nevertheless feature discrete rotational symmetry imposing restrictions on the maximal possible topological charge of such states, see works on lattices induced by different nondiffracting beams or on photonic crystals [27-33], as well as recent studies on vortex solitons in Penrose-like quasicrystals [34-36]. Very recently, thresholdless vortex solitons of topological origin were suggested and observed experimentally in aperiodic structures with disclinations [37,38] as well as in fractal waveguide arrays [39]. Vortex solitons can also exist in simple circular waveguide arrays, photonic crystal fibers with circular arrangement of guiding cores or multicore fibers [15,40-46], also belonging to a class of systems with discrete rotational symmetry [47,48]. In a narrow power range, they can be stabilized in parabolic waveguides [49]. The introduction of inhomogeneous gain and losses can introduce very unusual features into properties of vortex solitons [50,51], for example breaking the equivalence of states with opposite topological charges.

Recently, a new class of lattices in the form of star junctions (or networks) attracted considerable attention in optics and physics of topological systems. Such lattices represent several line “chains” or arrays of guiding elements with coupling between different arrays occurring only locally. The most remarkable feature of such lattices is that they can support localized modes that emerge only from local variation of the effective refractive index around coupling point. Theoretically star junctions have been studied mostly in discrete optical and matter-wave systems [52-57], while experimentally they were realized using locally coupled waveguide arrays that allowed to observe trivial-phase localized modes in the point, where arrays approached each other [58,59]. Localization occurs despite the fact that couplings constants between all nearest neighbors was the same, i.e. this is geometrical effect. Non-Hermitian generalizations of such systems have been considered too [60]. It has been shown that composition of star networks from topologically nontrivial chains dramatically increases the variety of localized linear states that they can support [61,62]. Nevertheless, so far only localized states with trivial phase distributions have been studied in star networks. The possibility of construction of star networks capable of supporting localized vortex states has never been illustrated and the transformation of such states under the action of nonlinearity has not been studied.

Here I propose star networks composed from several locally coupled line waveguide arrays that can support localized vortex states with different topological charges, increasing with increase of complexity of the network (number of rays in it). The first waveguide in each such line array resides on the ring of variable radius that can be changed to control the emergence of the localized modes arising from local couplings between different arrays brought in close proximity. The system can be also considered as a ring waveguide array coupled to several semi-infinite line waveguide arrays (i.e. as a system initially supporting only localized modes coupled to system supporting only continuum of extended states). Remarkably, localized vortex modes in this system can emerge with propagation constants not only above, but also below the band of the extended states. Such linear modes give rise to families of thresholdless vortex solitons, when focusing nonlin-earity of the medium is taken into account. Such solitons exhibit nontrivial stability properties, depending on the discrete rotational symmetry of the system, that are sharply distinct from stability properties of vortex solitons constructed on the "isolated" ring of waveguides.

## 2. Theoretical model and network configuration

We consider propagation of paraxial light beams in focusing cubic nonlinear medium with shallow transverse refractive index modulation defining waveguide star network. The propagation dynamics is governed by the nonlinear Schrödinger equation for the normalized amplitude $\psi$ of the light field:

$$i\frac{\partial\psi}{\partial z}=-\frac{1}{2}\left(\frac{\partial^2\psi}{\partial x^2}+\frac{\partial^2\psi}{\partial y^2}\right)-|\psi|^2\psi-\mathcal{R}(x,y)\psi. \quad (1)$$

Here I use for normalization standard soliton units assuming that star network, whose refractive index profile is described by the function $\mathcal{R}(x,y)$, is created using technology of femtosecond-laser writing in transparent nonlinear dielectric, such as fused silica, see review [63], as well as experiments [38,39]. In these units, transverse coordinates $(x,y)$ are normalized to the characteristic scale $r_0$ (for example, one can choose $r_0$=10 $\mu$m ); the propagation distance $z$ is normalized to diffraction length $kr_0^2$; $k=2\pi n/\lambda$ is the wavenumber; $n\approx 1.45$ is the unperturbed refractive index of the material; $\lambda=800$ nm is the typical working wavelength for observation of spatial solitons in such systems; the intensity $|\psi|^2$ corresponds to $I=n|\psi|^2/k^2r_0^2n_2$, where $n_2$ is the nonlinear refractive index of the material, in the case of fused silica $n_2\approx 2.7\times10^{-20}$ m$^2$/W.

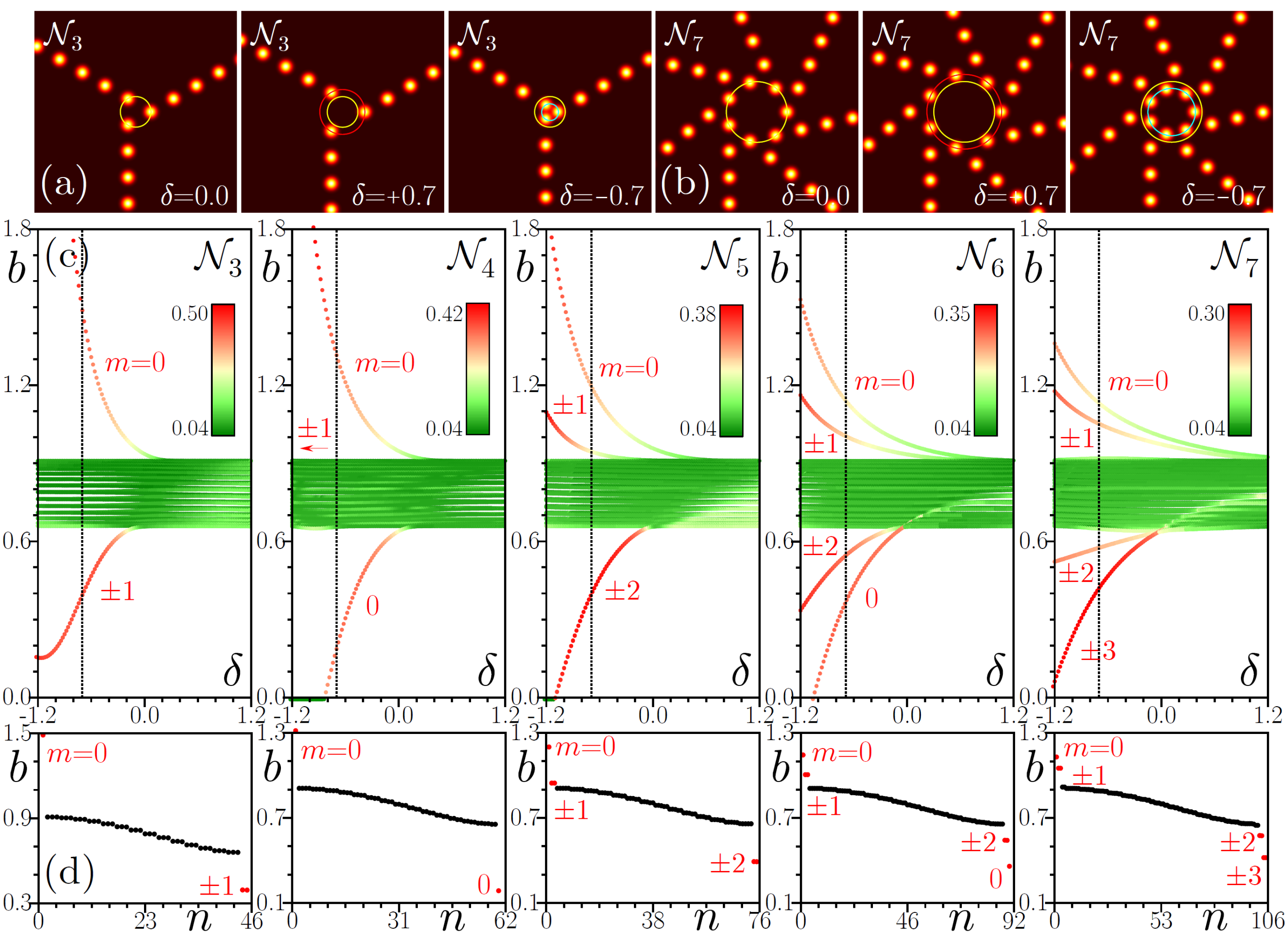

Fig. 1. Examples of the refractive index distributions $\mathcal{R}(x,y)$ in (a) $\mathcal{N}_3$ and (b) $\mathcal{N}_7$ star networks for different values of shift parameter $\delta$ indicated in each panel. At $\delta=0$ first waveguides in each ray are located on yellow circle; red and cyan circles show locations of such waveguides at $\delta>0$ and $\delta<0$, respectively. (c) Eigenvalues of all linear modes supported by various $\mathcal{N}_k$ star networks vs shift parameter $\delta$. Color coding indicates form-factors $\chi$ of linear modes, from low (green color, delocalized modes) to high (red color, localized modes) ones. Red labels near each curve detaching from bulk band indicate topological charges of localized vortices that can be constructed using combinations of respective linear modes. (d) Eigenvalues of all linear modes versus mode index $n$ at $\delta=-0.7$ [corresponding to dotted lines in panels (c)]. In all cases $p=5$, $d=2.6$, $a=0.5$.

Star networks considered here are constructed from several "line" waveguide arrays brought in close proximity [see Fig. 1(a) and 1(b)], where all Gaussian waveguides $\mathcal{Q}(x,y)=pe^{-(x^2+y^2)/a^2}$ forming the network $\mathcal{R}(x,y)=\sum_{l,m}\mathcal{Q}(x-x_m,y-y_l)$ are identical [here coordinates $(x_m,y_l)$ determine the positions of the waveguide centers]. Here $p$ is the waveguide depth and $a$ is its width. The spacing between waveguides in line arrays is fixed and is equal to $d$. The first waveguide in each line array is located on a ring with initial radius $r = d/(2\cos\varphi)$ [see the yellow ring in Fig. 1(a) and 1(b)], where $\varphi = \pi(\mathcal{N}_{\text{rays}} - 2)/\mathcal{N}_{\text{rays}}$, with $\mathcal{N}_{\text{rays}} = 3, 4...$ denoting the number of "rays" in star network. The angle $\varphi$ is the internal angle of polygon formed by waveguides on a ring and it is selected such that the side of this polygon is equal to waveguide spacing $d$ in each ray. In such network the spacing between all nearest waveguides in each ray and between first waveguides in different rays is the same and is equal to $d$. To introduce additional tunability into structure, the radius $r$ of the central waveguide ring can be changed to $r + \delta$ and increases for $\delta > 0$ [red ring in Fig. 1(a),1(b)] and decreases for $\delta < 0$ [cyan ring in Fig. 1(a),1(b)]. The shift parameter $\delta$ will be used as a main control parameter determining the appearance of vortex carrying states, as shown below. Further, the width $a = 0.5$ $(5\ \mu\text{m})$, spacing $d = 2.6$ $(26\ \mu\text{m})$, and depth $p = k^2 r_0^2 \delta n / n = 5$ $(\delta n \sim 5.6 \times 10^{-4})$ of the waveguides will be considered fixed. As one can see from Fig. 1 the constructed star networks feature discrete rotational symmetry $\mathcal{C}_n$, whose order is equal to the number of rays $\mathcal{N}_{\text{rays}}$ in structure. This has direct implications for the possible topological charges of vortex modes that can exist in such structures.

## 3. Linear vortex modes in star networks

To understand the properties of this system it is instructive to consider first the spectrum of its linear eigenmodes. At this step the focusing nonlinearity in Eq. (1) can be neglected, and one can search for linear modes in the form $\psi = w(x,y)\exp(ibz)$, where $w(x,y)$ is the complex function describing eigenmode profile and $b$ is its eigenvalue (propagation constant). This substitution into Eq. (1) yields linear eigenvalue problem:

$$bw = \frac{1}{2}\left(\frac{\partial^2 w}{\partial x^2} + \frac{\partial^2 w}{\partial y^2}\right) + \mathcal{R}(x,y)w, \qquad (2)$$

which was solved using plane-wave expansion method. In addition to calculation of all eigenmodes of the system $w_n$ with eigenvalues $b = b_n$, for each such mode the form-factor $\chi_n = (\iint |w_n|^4\, dxdy)^{1/2}$ was obtained (here each mode is normalized such that its power $U_n = \iint |w_n|^2\, dxdy = 1$). The form-factor is inversely proportional to the mode width, so that the values $\chi_n \sim 1$ imply strong localization, while $\chi_n \ll 1$ means that the mode with index $n$ is delocalized.

In Fig. 1(c) the eigenvalues of all modes of star networks with different progressively increasing number of rays $\mathcal{N}_{\text{rays}}$ are shown as the functions of the shift parameter $\delta$. In addition, color coding of dots indicating eigenvalues $b_n$ of different modes illustrate also the form-factor of this linear mode, with green color corresponding to delocalized modes, while red color corresponds to localized modes [see color-bars in different panels of Fig. 1(c)]. As one can see, the common feature for all star networks is the presence of the band of delocalized states (green band in all panels). States from this band extend along the entire rays of the structure. Remarkably, the decrease of shift parameter $\delta$ results in detaching from the band of delocalized states of localized (around central ring) linear eigenmodes with different topological charges $m$. It must be emphasized that the fundamental eigenmode with trivial phase structure (topological charge $m = 0$) exists in star networks even for zero shift parameter $\delta$. Such modes always appear above the band of delocalized states. These are classical eigenmodes resulting from local increase of the effective refractive index due to proximity of different rays of the structure and they are similar to modes on topological array defects encountered in [58]. The examples of such modes are presented in the left outermost panels in Fig. 2(a) and 2(b). Notice that the separation between propagation constant of such modes and the band of extended states increases with increase of the number of rays $\mathcal{N}_{\text{rays}}$ in the array.

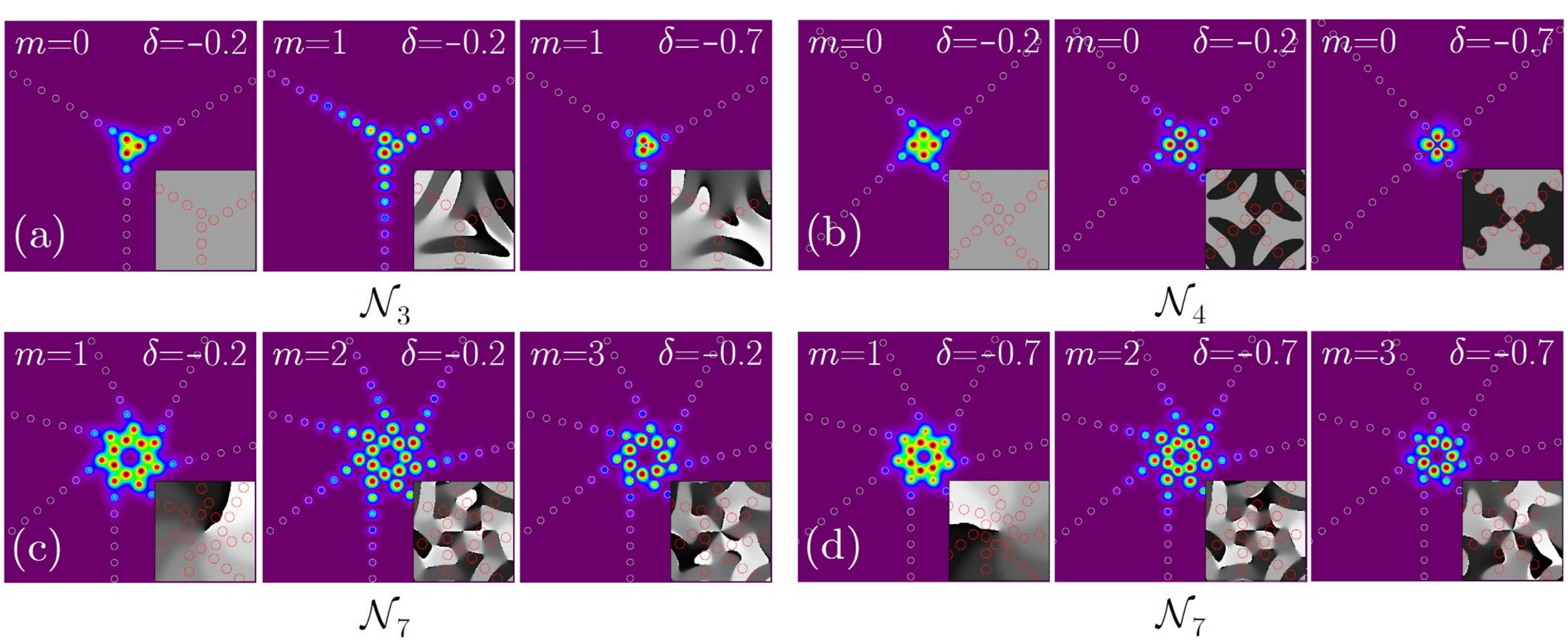

Fig. 2. Examples of field modulus and phase (insets) distributions in linear modes with different topological charges $m$ supported by star networks $\mathcal{N}_3$ (a), $\mathcal{N}_4$ (b), and $\mathcal{N}_7$ (c),(d) at different values of shift parameter $\delta$ indicated in each panel. Gray and red circles indicate waveguide positions. Field modulus distributions are shown within $x, y \in [-20, +20]$ window, while phase distributions in the insets are shown within $x, y \in [-8, +8]$ window.

The remarkable feature of the structure proposed here is that in addition to these simplest modes, the negative values of the shift parameter $\delta$ result in detaching from extended band of linear modes carrying vorticity $(m \neq 0)$. For example, for $\mathcal{N}_3$ structure such linear modes carry topological charges $m = \pm 1$ [vortex states always emerge in degenerate pairs with opposite values of the topological charge because the function $\mathcal{R}(x,y)$ describing array is real, hence in Fig. 1(c) the charge near corresponding curves is written as $\pm m$ ] and this is the maximal value of topological charge allowed by the discrete rotational symmetry of this star network. Surprisingly, these vortex modes in $\mathcal{N}_3$ network appear below the band of the extended states. As a result, the field in such modes changes its sign between neighboring waveguides in each ray leading to very unusual phase distributions [see examples in Fig. 2(a),2(c) and 2(d), middle panels]. This is an indication of Bragg-like mechanism of localization for all modes appearing below the band of extended states. In contrast, in all modes emerging above the band of extended states, the field does not change its sign along the rays [Fig. 2(a),2(b), left panels], meaning that such modes emerge due to local increase of the effective refractive index around the central ring. Increasing the number of rays in network by $1$ adds one linear localized mode, whose total number is always equal to $\mathcal{N}_{\text{rays}}$ (for $\mathcal{N}_4$ a pair of $m=\pm 1$ states appear at too large negative shifts $\delta$, when waveguides start overlapping, and is not visible on the scale of the figure). While the mode with the largest eigenvalue always has trivial phase $(m=0)$, the lowest mode can be vortical (for odd number or rays) or multipole (for even number of rays). The degree of localization of all modes of star network depend on how large is the deviation of propagation constant of a given mode from the band of extended states. The example is shown in Fig. 2(c) for $\mathcal{N}_7$ , where at $\delta=-0.2$ the mode with topological charge $m=2$ has propagation constant closer to the band of extended states than $m=3$ mode and on this reason it features substantially weaker localization along the rays of the network in comparison with latter mode. The localization of all eigenmodes increases with decrease of the shift parameter $\delta$ , as one can see from comparison of mode profiles with the same charge $m$ in Fig. 2(c) at $\delta=-0.2$ and Fig. 2(d) at $\delta=-0.7$ . Interestingly, for large number of rays in the network, when coupling between them becomes stronger due to tighter packing, localized vortex-carrying modes may emerge even at sufficiently large positive shifts $\delta$ . The largest network that I consider here contains $\mathcal{N}_7$ rays, and it can supports vortex modes with topological charges up to $m=\pm 3$ [Fig. 1(c)], also consistent with limitations imposed by the discrete rotational symmetry, see details in [32,33]. Figure 1(d) shows propagation constants of all eigenmodes versus mode index $n$ at fixed value of shift parameter $\delta=-0.7$ corresponding to the dashed lines in Fig. 1(c). These plots clearly illustrate that vortex-carrying states appear in degenerate pairs.

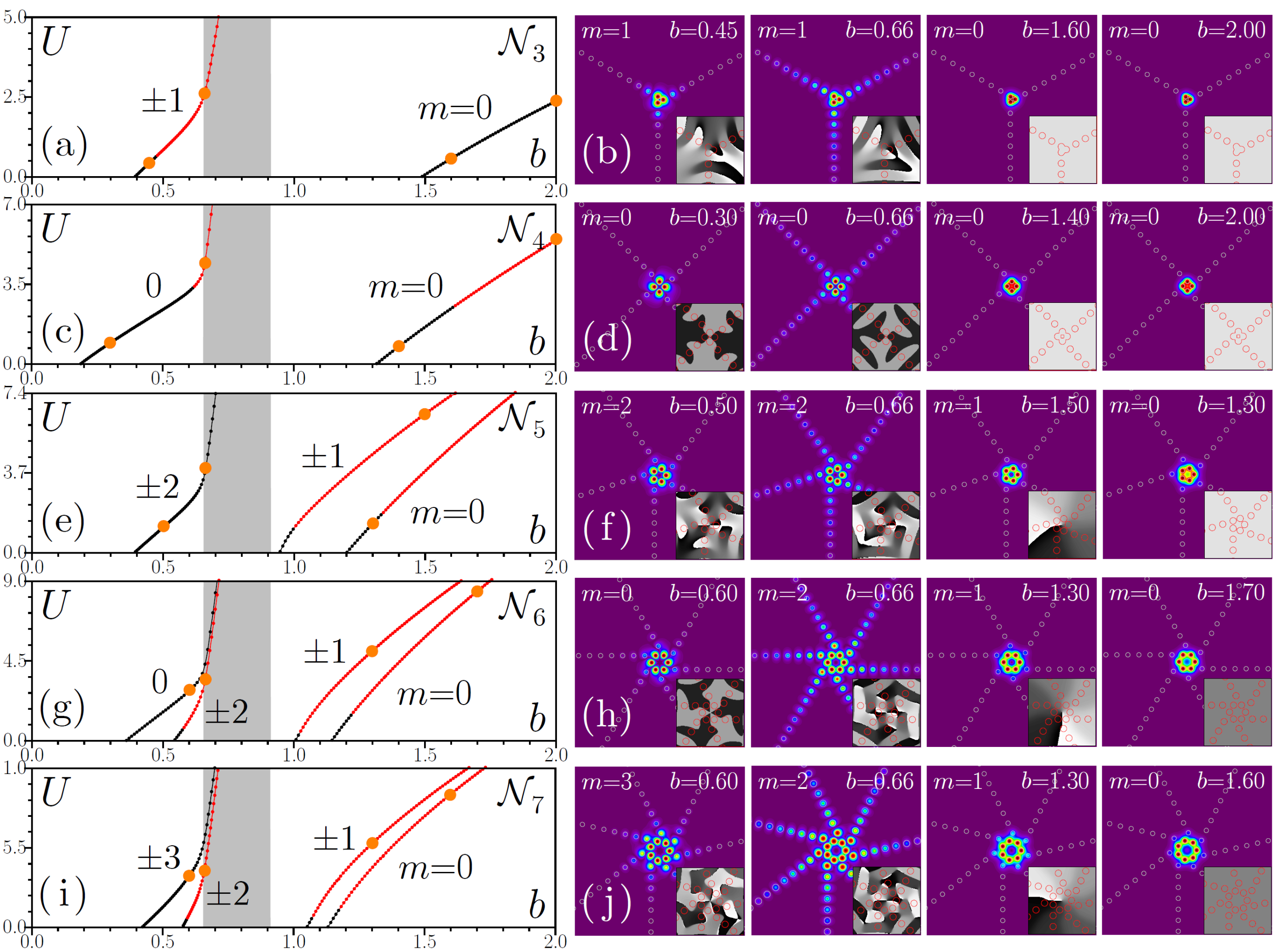

Fig. 3. $U(b)$ dependencies illustrating soliton families with different topological charges $m$ in star networks and corresponding field modulus and phase (insets) distributions for networks $\mathcal{N}_3$ (a),(b), $\mathcal{N}_4$ (c),(d), $\mathcal{N}_5$ (e),(f), $\mathcal{N}_6$ (g),(h), and $\mathcal{N}_7$ (i),(j). Gray areas show band of extended linear modes. Soliton profiles shown at the right corresponding to the orange dots in $U(b)$ dependencies, their propagation constants are indicated on each profile. Field modulus distributions are shown within $x,y\in[-20,+20]$ window, while phase distributions in the insets are shown within $x,y\in[-8,+8]$ window. Black branches correspond to stable solitons, while red branches are unstable. In all cases $\delta=-0.7$ .

## 4. Solitons in star networks and their stability

Next, focusing nonlinearity of the medium is taken into account. In this case, each localized linear mode gives rise to a family of solitons bifurcating from them. Such solitons can be found from Eq. (1) in the form $\psi = w(x,y)\exp(ibz)$, where now propagation constant dictates the shape of soliton state $w(x,y)$. The families of solitons can be characterized by the dependence of their power $U = \iint |w|^2 dxdy$ on propagation constant $b$. Such dependencies are presented in the left column of Fig. 3 for networks from $\mathcal{N}_3$ to $\mathcal{N}_7$ for representative value of shift parameter $\delta = -0.7$ (the results for other $\delta$ values are qualitatively similar). Because they bifurcate from the linear localized modes, solitons in star network are thresholdless, i.e. their power $U$ vanishes when propagation constant $b$ approaches the eigenvalue of respective linear mode. Power monotonically increases with increase of $b$, but solitons bifurcating from states above and below the band of extended states (gray regions in plots in left column of Fig. 3) exhibit completely different localization properties. For solitons emerging above the band of extended states localization progressively increases with increase of $b$ [see, for example, the distributions for $m=0$ states in Fig. 3(b) and 3(d)]. In contrast, solitons bifurcating from eigenmodes below bulk band broaden as their propagation constants approach the bulk band and eventually they delocalize due to coupling with extended states. Delocalization is manifested in considerable expansion of the mode along each ray of the network [see examples in second panels of Fig. 3(d),3(f),3(h) and 3(j)]. Such delocalization occurs both for multipoles with $m=0$ and for vortex-carrying modes with $m \neq 0$. Comparing phase structure of vortex solitons above [Fig. 3(j), third panel] and below [Fig. 3(j), first and second panels] the band of extended states, one can see that the former do not exhibit radial phase oscillations, while in latter state the phase changes its sign between waveguides in each ray (in addition to overall imposed vorticity).

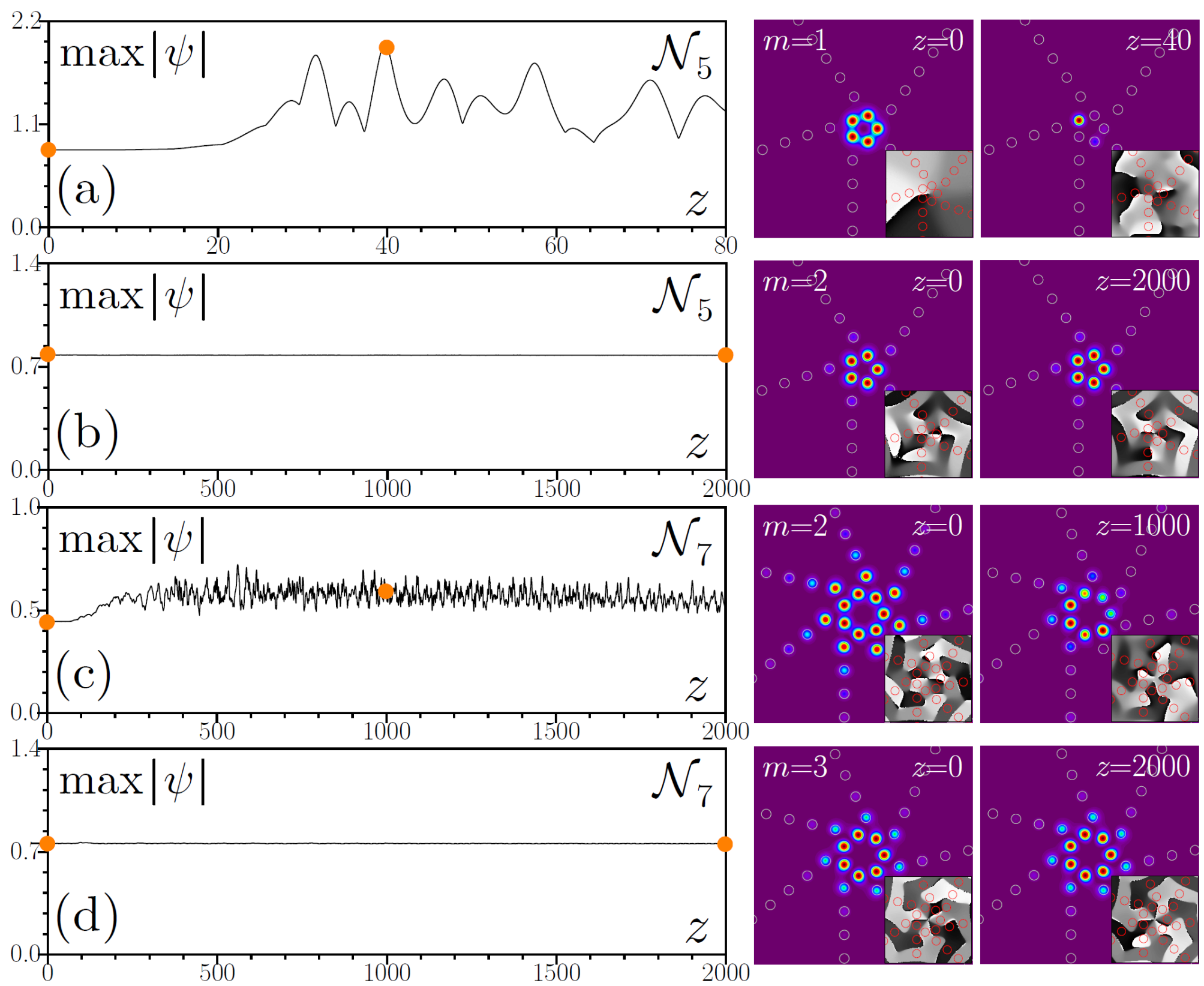

Fig. 4. Example of instability development of vortex soliton with $b=1.20$, $m=1$ in $\mathcal{N}_5$ network (a), stable propagation of soliton with $b=0.64$, $m=2$ in $\mathcal{N}_5$ network (b), instability of soliton with $b=0.65$, $m=2$ in $\mathcal{N}_7$ network (c), and stable dynamics of $b=0.64$, $m=3$ state in $\mathcal{N}_7$ network (d). Field modulus and phase (insets) distributions at different distances correspond to the orange dots in peak amplitude $\max|\psi|$ vs $z$ dependencies. Field modulus distributions are shown within $x,y \in [-12,+12]$ window, while phase distributions in the insets are shown within $x,y \in [-8,+8]$ window. In all cases $\delta = -0.7$.

Solitons in star networks exhibit unusual stability properties that differ from stability of vortex solitons on a ring of waveguides [33]. Stability was tested by modeling direct evolution in the frames of Eq. (1) of slightly perturbed solitons $\psi|_{z=0} = w(x,y)[1+\rho(x,y)]$, where $\rho(x,y)$ is the function describing broadband complex noise with amplitude up to $0.05$. Such perturbed solutions were propagate to large distances $z$ to detect the presence of even weak instabilities. The results of stability analysis are summarized in $U(b)$ dependencies in the left column of Fig. 3, where stable soliton branches are shown black, while unstable branches are shown red. All solitons are stable at least in small vicinity of the bifurcation point from linear eigenmodes. For $m=0$ solitons above the band of extended states azimuthal modulation instability can develop only when the power of such solitons exceeds certain critical value and this valued decreases with increase of the number of rays in structure [c.f. broad stability domain for $\mathcal{N}_3$ network in Fig. 3(a) and narrow stability domain for $\mathcal{N}_7$ network in Fig. 3(i)]. Notice that instability of such states is usually accompanied by the bifurcation from corresponding families with identical spots on a ring, a family of asymmetric solitons (they are not shown in Fig. 3). The stability of family with identical spots is lost exactly in the bifurcation point. A general observation is that besides narrow stability domains present for all soliton branches near the bifurcation points from linear states, the most stable branches are those that bifurcate from linear modes with the lowest propagation constants (i.e. left outermost branches in Fig. 3). For networks with odd number of rays, like $\mathcal{N}_5$ [Fig. 3(e)] and $\mathcal{N}_7$ [Fig. 3(i)], these are the branches of the vortex solitons with $m=\pm 2$ and $m=\pm 3$, respectively. The exception is $\mathcal{N}_3$ network, where the lowest branch with $m=\pm 1$ appears to be stable only near the bifurcation point. For networks with even number of rays, like $\mathcal{N}_4$ [Fig. 3(c)] and $\mathcal{N}_6$ [Fig. 3(g)], the stable branches are represented by multipoles with $m=0$. Remarkably, some of the left outermost branches remain stable even when they enter into the band of extended states and corresponding solitons become delocalized [see Fig. 3(c),3(e),3(g) and 3(i)]. The presence of stability islands near the bifurcation points and stability mostly for left outermost branches is what distinguishes this system from conventional ring waveguide arrays [33].

The examples of stable propagation of different vortex solitons and instability development are shown in Fig. 4. Unstable vortex solitons emerging above the band of extended states show standard azimuthal modulation instability, after development of which light typically concentrates in one of the waveguides on the central ring of the structure, while vorticity is typically lost [see example for $m=1$ vortex soliton in $\mathcal{N}_5$ network in Fig. 4(a)]. The example of stable propagation of the $m=2$ vortex soliton in the same network is shown in Fig. 4(b). This stable solution maintains its structure over the entire propagation distance. Solitons, bifurcating from linear modes below the band of extended states may feature weak oscillatory instabilities typical for gap lattice solitons. The example of such instability is illustrated in Fig. 4(c) for $m=2$ vortex soliton in $\mathcal{N}_7$ network. As a result of this instability development, soliton typically acquires azimuthal modulation, but still covers multiple waveguides on the central ring. Its phase structure is lost in the course of propagation. Finally, the example of stable evolution of the $m=3$ vortex soliton in $\mathcal{N}_7$ network is shown in Fig. 4(d).

## 5. Conclusions

Summarizing, it is shown that star networks can support rich families of vortex solitons. Their complexity and available topological charges increase with increase of the number of rays in network, i.e. with increase of its discrete rotational symmetry. The characteristic feature of such systems is the presence of the band of extended states, because the network is formed by semi-infinite line waveguide arrays. The interactions of solitons with this band lead to unusual localization and stability properties. The results reported here can be potentially extended to networks composed from topological chains, or several locally coupled higher-order topological structures that may considerably enrich the spectrum of linear and nonlinear states in such systems.

**CRediT authorship contribution statement**

Y.V.K.: Writing – review & editing, Validation.

**Funding**

Y.V.K. acknowledges support from the research project FFUU-2024-0003 of the Institute of Spectroscopy of the Russian Academy of Sciences.

**Declaration of competing interest**

The author declares that he has no known competing financial interests or personal relationships that could have appeared to influence the work reported in this paper.

**Data Availability**

Data underlying the results presented in this paper may be obtained from the author upon reasonable request.